\documentclass[runningheads,a4paper]{llncs}

\makeatletter
\newif\if@restonecol
\makeatother

\usepackage[linesnumbered,ruled,vlined]{algorithm2e}%[ruled,vlined]{
\usepackage{algpseudocode}
\usepackage{amsmath}
\usepackage{amssymb}
\usepackage{graphicx}
\usepackage[misc]{ifsym}
\usepackage{bbding}
\usepackage{setspace}
\usepackage{amsmath}
\usepackage{mathrsfs}
\usepackage{booktabs}
\usepackage{tabularx}
\usepackage{color}

\usepackage{url}
\urldef{\mails}\path|{liguangli, liulei, wangxueying, dongxiao, zhaopeng, fxb}@ict.ac.cn|

\newcommand{\keywords}[1]{\par\addvspace\baselineskip
\noindent\keywordname\enspace\ignorespaces#1}

\begin{document}

\mainmatter  % start of an individual contribution

%\title{Auto-Tuning Neural Network Quantization Framework for Collaborative Inference \\Between the Cloud and Edge}

\title{Auto-Tuning Neural Network Quantization Framework for Collaborative Inference \\Between the Cloud and Edge\footnote{Published at ICANN 2018}}
\titlerunning{Auto-Tuning Neural Network Quantization Framework}
\author{Guangli Li\inst{1,2} \and Lei Liu \inst{1(}\Envelope\inst{)} \and
Xueying Wang\inst{1,2} \and Xiao Dong\inst{1,2} \and \\Peng Zhao\inst{1,2} \and Xiaobing Feng\inst{1}}
\authorrunning{Guangli Li, Lei Liu, Xueying Wang, et al.}

\institute{State Key Laboratory of Computer Architecture, \\Institute of Computing Technology, Chinese Academy of Sciences, Beijing, China \and
University of Chinese Academy of Sciences, Beijing, China\\
\mails}


\toctitle{Quantization Framework}
\tocauthor{Guangli Li, Lei Liu, Xueying Wang, et al.}
\maketitle

\begin{abstract}

Recently, deep neural networks (DNNs) have been widely applied in mobile intelligent applications.
The inference for the DNNs is usually performed in the cloud.
However, it leads to a large overhead of transmitting data via wireless network.
In this paper, we demonstrate the advantages of the cloud-edge collaborative inference with quantization.
By analyzing the characteristics of layers in DNNs,
an auto-tuning neural network quantization framework
for collaborative inference is proposed.
We study the effectiveness of mixed-precision collaborative inference
of state-of-the-art DNNs by using ImageNet dataset.
The experimental results show  that our framework can generate reasonable network partitions and reduce the storage on mobile devices with trivial loss of accuracy.

\end{abstract}

\section{Introduction}

In recent years, deep neural networks (DNNs) \cite{lecun2015deep} are widely
used and show impressive performance in various fields including computer vision \cite{krizhevsky2012imagenet}, speech recongnition\cite{hinton2012deep}, natural language processing \cite{mikolov2013distributed}, etc.
As the neural network architectures become more complex and deeper --- from LeNet \cite{Lecun1998Gradient} (5 layers) to ResNet \cite{He2015Deep} (152 layers),
the storage and computation of the model is increasing.
In other words, it leads to more resource requirements for network training and inference.
The large size of DNN models limits the applicability of the network inference on mobile edge devices. Therefore, most of artificial intelligence (AI) applications on mobile devices send input data of DNN to cloud servers, and the procedure of network inference is executed in the cloud only.
However, the cloud-only inference has some assignable weaknesses:
1) transmission overhead: it leads to a large overhead of uploading data especially when the mobile edge devices are in the low-bandwidth wireless environments.
2) privacy disclosure: sometimes, personal data, e.g. one's photos and videos, are not allowed to send to the cloud servers directly.

Today's mobile devices, such as Apple's iPhone and NVIDIA's Jetson TX2, have more powerful computability and larger memory.
In addition, many neural network quantization methods \cite{Cheng2018A,Cheng2017Quantized,Gong2014Compressing,Zhou2017Incremental,Zhou2016DoReFa} have been proposed for reducing the resource consumption of DNNs.
By using quantization, the data of a network can be represented by low-precision values, e.g. INT8 (8-bit integer).
On the one hand, low-precision data reduces storage of DNNs
and enables network models to be stored on the mobile edge device with limited resources.
On the other hand, with the use of high-performance libraries for low-precision computing \cite{google-gemm,tensorrt}, the speed of the network inference will be improved.
This makes it possible to perform some or all parts of neural network inference on mobile devices and leads to a new inference mode: cloud-edge collaborative inference.

\begin{figure}
\centering
\includegraphics[height=6.2cm]{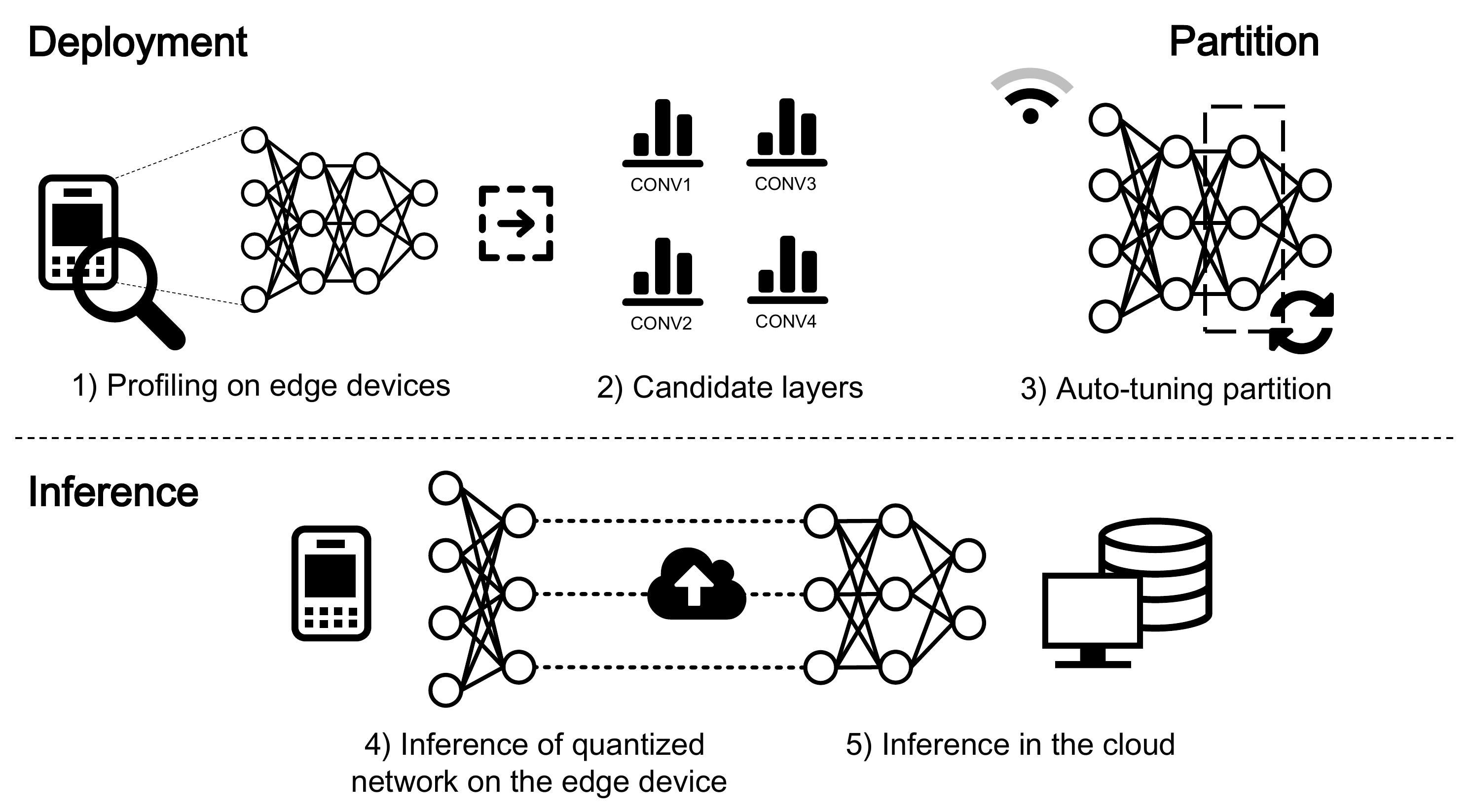}
\caption{Overview of auto-tuning framework}
\label{fig:overview}
\end{figure}

In this paper, we propose an auto-tuning neural network quantization framework as shown in Fig. \ref{fig:overview}.
During deployment, the framework profiles the operators of DNNs on edge devices
and generates the candidate layers as partition points.
When the neural network is ready to be used, the framework starts auto-tuning for network partition.
In the time of inference, the first part of the network is quantized and executed on the edge devices,
and the second part of the network is executed in the cloud servers.
On the edge, we use quantized neural network to reduce storage and computation.
In the cloud, we use original full-precision network to achieve high accuracy.

In the collaborative inference, quantized neural networks can reduce the storage of models.
Intermediate results of quantized networks are also low-precision data,
which can reduce data communication between cloud and edge.
So user's mobile device could transmit less data when using AI applications.
Additionally, transmitting intermediate result data, rather than the original input data, can protect personal information.
In realistic scenarios, the process of analysis and testing is tedious and time-consuming.
It's unfriendly for a program developer to test and decide how to partition the network.
Our automatic tuning framework will help developers find the most reasonable partition of a DNN. The contributions of this paper are summarized as follows:
\begin{itemize}
\item[$\bullet$] \emph{Analysis of DNN partition points} --
We analyze the structures of deep neural networks
and show which layers are reasonable partition points.
Based on the analysis, we could generate candidate layers
as partition points of a specific neural network.
(Section 2.2)

\item[$\bullet$] \emph{Auto-tuning quantization framework for collaborative inference} --
We develop an auto-tuning neural network quantization framework
for collaborative inference between cloud and edge.
The framework quantizes neural networks according to the candidate partition points
and provides an optimal mixed-precision partition for cloud-edge inference by auto-tuning.
(Section 2.3)

\item[$\bullet$] \emph{Experimental study} --
We show the performance of collaborative inference of state-of-the-art DNNs by using ImageNet dataset.
The framework generates reasonable network partitions and reduces the storage of inference on mobile devices with trivial loss of accuracy.
(Section 3)

\end{itemize}

\section{Auto-Tuning Quantization Framework}

In this section, we present our auto-tuning neural network quantization framework.
Firstly,  we briefly introduce neural network quantization.
Secondly, we analyze the structures of the state-of-the-art DNNs.
Finally, we describe the auto-tuning partition algorithm.

\subsection{Neural Network Quantization}

In order to accelerate inference and compress the size of DNN models,
many network quantization methods are proposed.
Some studies focus on scalar and vector quantization \cite{Cheng2017Quantized,Gong2014Compressing},
while others center on fixed-point quantization \cite{Zhou2017Incremental,Zhou2016DoReFa}.
In this paper, we are mainly interested in scalar quantization of INT8, which is supported by many advanced computing libraries such as Google's gemmlowp \cite{google-gemm} and NVIDIA's cuDNN \cite{tensorrt}.
In general, an operator computation of scalar quantized neural networks can be summarized as follows:

\begin{itemize}

\item[$\bullet$] Off-line Quantization \\
Step 1.  Find quantization thresholds ($T_{min}$ and $T_{max}$) for calculating scale factors of $Input$, $Weights$ and $Output$;

Step 2.  Quantize $Input$ and $Weights$ according to the following formula: \\
\begin{equation}
\label{eq1}
Data_Q(x)=\left\{
\begin{aligned}
%\frac{Range_{LP}}{|T_{max}-T_{min}|} \times (Data(x)-T_{min}) \qquad & {x \in (T_{min}, T_{max})} \\
\frac{Data(x)-T_{min}}{|T_{max}-T_{min}|} \times Range_{LP} \qquad & {x \in (T_{min}, T_{max})} \\
\Vert{V_{low-precision}}\Vert_{\infty}  \qquad & {x \ge T_{max}}\\
\Vert{V_{low-precision}}\Vert_{-\infty}  \qquad & {x \le T_{min}}
\end{aligned}
\right.
\end{equation}
where: ${Range_{LP}}$ is the range of low-precision values (e.g. 255 for INT8), $V_{low-precision}$ is the set of low-precision values, $Data(x)$ is the original value, $Data_Q(x)$ is the quantized value. \\

\item[$\bullet$] On-device Computation \\
Step 1.  $Output_Q$ = Operator($Input_Q$, $Weights_Q$);

Step 2.  Dequantize $Output_Q$ according to the following formula:\\
\begin{equation}
\label{equ2}
Output = \frac{|T_{max}-T_{min}|}{Range_{LP}} \times Output_Q(x) + T_{min}
\end{equation}

Step 3.  $Output$ = ActivationFunction($Output$);

Step 4.  Quantize $Output$ as $Input_{Next}$ according to Formula. \ref{eq1}.

\end{itemize}

\subsection{Candidate Network Partition Points}

In general, a deep neural network contains many kinds of layers such as
convolution layers, fully-connected layers and activation layers.
We analyze the characteristics of different network layers
and decide how to select candidate layers as reasonable partition points.
The set of candidate layers, $Rule=\{L_1,L_2,\ldots ,L_n\}$,
is based on the results of the following analysis.

\begin{figure}[!h]
\centering
\includegraphics[height=6.2cm]{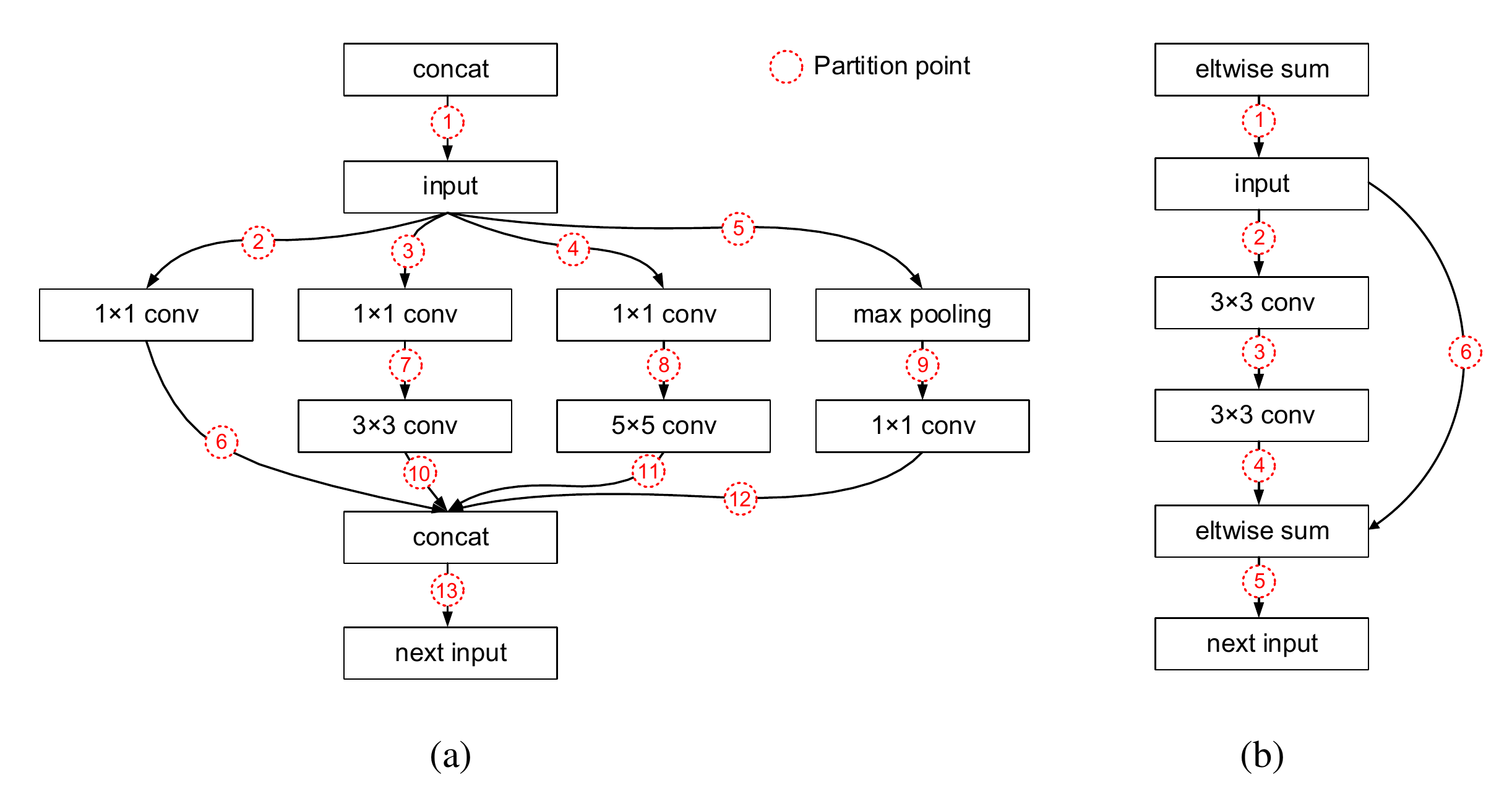}
\caption{Partition points of DNNs}
\label{fig:inception_resnet}
\end{figure}

\subsubsection{Layers in Inception Networks}

\begin{table}[htbp]
\centering
\caption{Analysis of inception}
\label{table1}
\begin{tabular}{|l|c|c|c|}
\hline
\textbf{\begin{tabular}[c]{@{}l@{}}Partition\\ points\end{tabular}}              & \multicolumn{1}{l|}{\textbf{\begin{tabular}[c]{@{}l@{}}Brother branch \\ exists?\end{tabular}}} & \multicolumn{1}{l|}{\textbf{\begin{tabular}[c]{@{}l@{}}Inference mode of \\the brother branch\end{tabular}}} & \multicolumn{1}{l|}{\textbf{\begin{tabular}[c]{@{}l@{}}Data Transmission\end{tabular}}} \\ \hline
1, 13  & No & / & INT8 $\times$ 1 \\ \hline
\begin{tabular}[c]{@{}l@{}}2, 3, 4, 5\\ 7, 8, 9\\ 6, 10, 11, 12\end{tabular} & Yes                     & Mobile Edge & INT8 $\times$ 4 \\ \hline
\begin{tabular}[c]{@{}l@{}}2, 3, 4, 5\\ 7, 8, 9\\ 6, 10, 11, 12\end{tabular} & Yes                    & Cloud & INT8 $\times$ 1 + FP32 $\times$ 1 \\ \hline
\end{tabular}
\end{table}

Inception is a structure that contains branches,
and these branches are executed in parallel
and their results are merged into a network layer (e.g. concat layer).
Fig. \ref{fig:inception_resnet}(a) is an example of inception from GoogLeNet \cite{Szegedy2015Going}.
As shown, the inception contains 13 possible partition points.
If we try all the partition points, it will take a lot of time.
We divide these partition points into two groups
according to whether they have at least a brother branch
(separate from the same layer and merge in the same layer).
The results of the analysis are shown in Table. \ref{table1}.
When a partition point has no brother branch (e.g. 1 and 13),
the output of the sub-network on edge devices contains only 1 $\times$ INT8 Blob
(4D array for storing data).
When a partition point has a brother branch, there are two cases:
1) its brother branch runs on the edge devices,
and the sub-network output contains 4 $\times$ INT8 Blobs;
2) its brother branch runs in the cloud,
and the sub-network output contains 1 $\times$ INT8 Blob and 1 $\times$ FP32 Blob.
The transmission data in first group
is smaller than it is in the second group.
Therefore, if a network layer in inception has a brother branch,
the framework will not choose it as a candidate layer.

\subsubsection{Layers in Residual Networks}

\setlength{\abovecaptionskip}{0.cm}
\begin{table}[htbp]
\centering
\caption{Analysis of residual network}
\label{table2}
\begin{tabular}{|l|c|c|}
\hline
\textbf{\begin{tabular}[c]{@{}l@{}}Partition \\points\end{tabular}} & \multicolumn{1}{l|}{\textbf{\begin{tabular}[c]{@{}l@{}}Shortcut connection exists?\end{tabular}}} & \multicolumn{1}{l|}{\textbf{Data Transmission}} \\ \hline
1, 5 & No  & INT8 $\times$ 1 \\ \hline
2, 3, 4, 6 & Yes & INT8 $\times$ 1 + FP32 $\times$ 1 \\ \hline
\end{tabular}
\end{table}
\setlength{\belowcaptionskip}{-0.cm}

There are many shortcut connections in the residual network \cite{He2015Deep}.
Fig. \ref{fig:inception_resnet}(b) shows an example of a residual block which contains a shortcut connection.
There are 6 possible partition points in this example.
According to whether the shortcut connection of a partition points exists,
we divide these partition points into two groups.
When a partition point has no shortcut connection (e.g. 1 and 5),
the output of the sub-network on edge devices contains only 1 $\times$ INT8 Blob.
Otherwise, the output of the sub-network contains 1 $\times$ INT8 Blob and 1 $\times$ FP32 Blob.
Table. \ref{table2} shows the analysis result.
Therefore, the network layers with shortcut connections are not reasonable candidate layers.

\subsubsection{Non-parametric Layers}
%\subsubsection{Non-parametric Layers and Plugin Layers}
Non-parametric layers, such as ReLU and pooling, have no parameters,
so they require almost no memory storage.
In addition, the computation of the non-parametric layers accounts
for a very small proportion of the total network computation.
Therefore, our framework merges the non-parametric layers into the nearest previous parametric layers,
i.e. these non-parametric layers will not be used as candidate layers.
% Plugin layers (e.g. RPN layers\cite{rcnn}) are atypical network layers designed by users
% which often not support low-precision computing.
% We will not consider plugin layers as candidate layers.

\subsection{Auto-Tuning Partition}

According to the candidate rule $Rule$,
the framework performs auto-tuning partition for cloud-edge collaborative inference,
as described in Algorithm. \ref{algo1}.
The input of the algorithm contains candidate layer rules and a neural network.
Firstly, candidate rules are used to select candidate partition points in the neural network (lines 1-2).
Secondly, all candidate partition networks are tested, and the information of performance is recorded in $P$ (lines 3-9).
The function of $PredictPerformance$ can predict the performance of collaborative inference based on the results of off-line profiling.
Finally, we find the best partition point in $P$ for collaborative inference of mixed-precision neural network (lines 10-14).

\begin{algorithm}[!h]
  \label{algo1}
  \caption{Auto-Tuning Partition}
  \KwIn{candidate rules $Rule$, neural network $Net={\{L_1,L_2,\ldots ,L_n\}}$}
  \KwOut{optimize partition $p_{best}$}
  $P \gets \Phi$; $p_{best} \gets null$\;
  $Candidate \gets \{L_i | L_i \in Rule\}$\;
  \For{$L_i$ in $Candidate$}
  {
    $Net_{edge} \gets Net.Split(First, L_i)$\;
    $Net_{Cloud} \gets Net.Split(L_i+1, Last)$\;
    $Engine_{Edge} \gets Net_{Edge}(DataType_{<INT8>})$\;
    $Engine_{Cloud} \gets Net_{Cloud}(DataType_{<FP32>})$\;
    $(L_i, info) \gets PredictPerformance(Engine_{Edge}, Engine_{Cloud})$\;
    $P \gets P \cup (L_i, info)$\;
  }
  $Env = GetEnvironment(Device_{Edge})$\;
  \For{$p_i$ in $P$}
  {
  	\If{$Env(p_i)$ is better than $Env(p_{best})$}
    {
    	$p_{best} \gets p_i$\;
    }
  }
  return $p_{best}$\;
\end{algorithm}

\section{Experiments}

In this section, we use ImageNet \cite{DengDSLL009} dataset to test the collaborative inference of DNNs \cite{He2015Deep,krizhevsky2012imagenet,Simonyan2014Very,Szegedy2015Going} and show results of our auto-tuning framework.
We illustrate the most reasonable partition for each neural network.
The inference of the edge performs on a mobile platform -- NVIDIA Jetson TX2 (NVIDIA's latest mobile SoC) -- with 4 $\times$ ARM Cortex-A57 CPUs and 2 $\times$ Denver CPUs, 8G of RAM.
The inference of the cloud performs on a server with Intel Core-i7 CPU, NVIDIA TITAN Xp GPU, 16G of RAM.
We use Caffe \cite{Jia2014Caffe} with cuDNN (version 7.0.5) on the GPU of cloud servers.
We use gemmlowp's \cite{google-gemm} implementation on the CPU of the edge devices.

\subsection{Experimental Results}

\begin{figure}[!h]
\centering
\includegraphics[width=12cm]{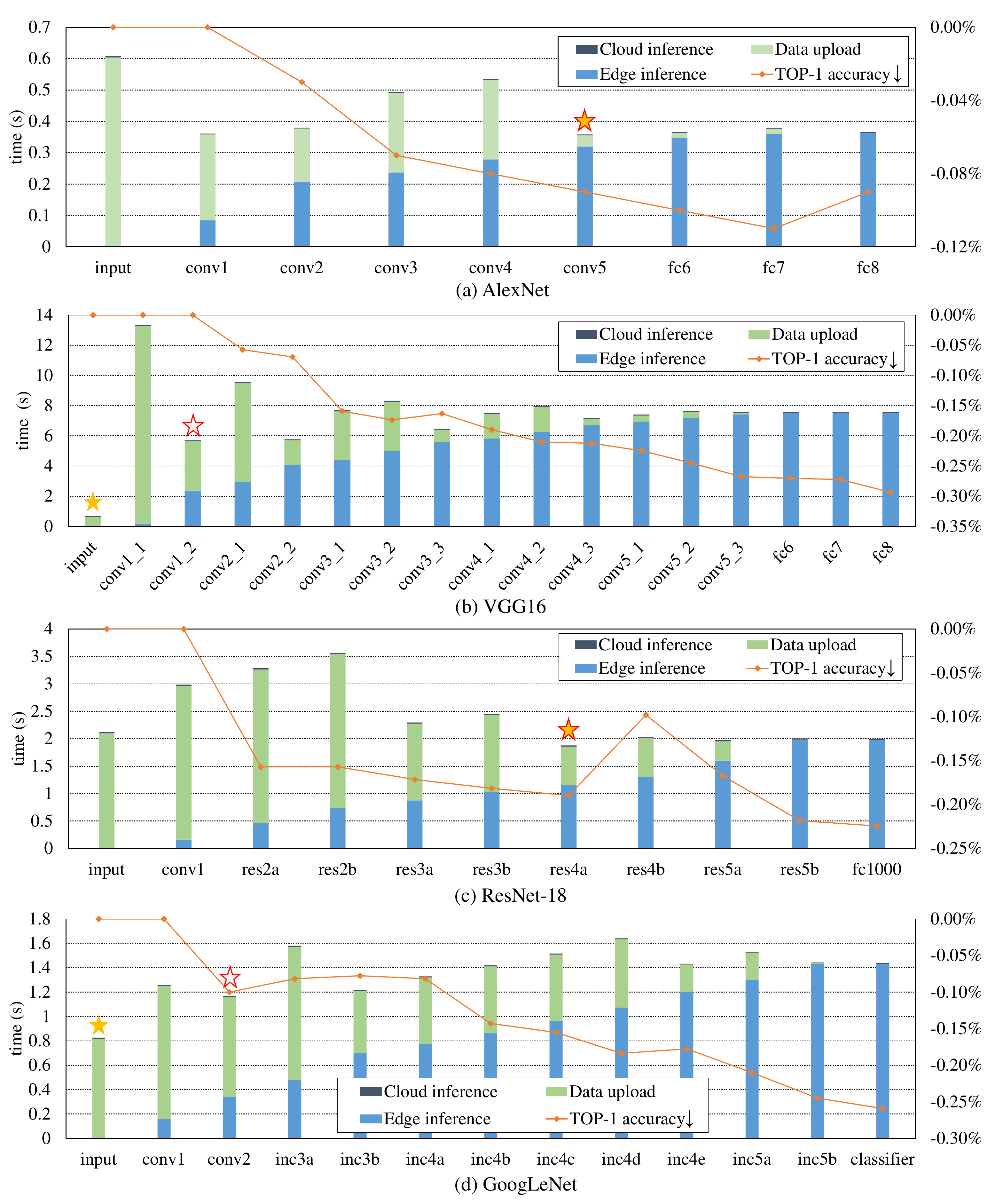}
\caption{Performance of each DNN partition}
\label{p_results}
\end{figure}

\begin{table}[!h]
\centering
\caption{Experimental results of our framework}
\label{results}
\begin{tabular}{|l|r|r|r|r|}
\hline
\textbf{Neural network} & \textbf{AlexNet} & \textbf{VGG16} & \textbf{ResNet-18} & \textbf{GoogLeNet} \\ \hline
\textbf{Wireless upload (KB/s)} & 250 & 240 & 70 & 180 \\ \hline
\textbf{Best partition point} & conv5 & conv1\_2 & res4a & conv2 \\ \hline
%\textbf{Fastest partition point} & conv5 & input & res4a & input \\ \hline
\textbf{Inference time (s)} & 0.36 & 5.65 & 1.86 & 1.16 \\ \hline
\textbf{Speed-up} & 1.7$\times$ & \textless{}1$\times$ & 1.13$\times$ & \textless{}1$\times$ \\ \hline
\textbf{Model download (KB)} & 2278 & 38 & 1569 & 121 \\ \hline
\textbf{Model storage reduction} & 96.17\% & 99.97\% & 85.63\% & 98.22\% \\ \hline
\textbf{TOP-1 accuracy$\downarrow$} & -0.09\% & 0.00\% & -0.19\% & -0.10\% \\ \hline
\end{tabular}
\end{table}

Table. \ref{results} summarizes the results of our framework.
We tested AlexNet, VGG16, ResNet-18 and GoogLeNet in different wireless network environments.
For each neural network, the framework gives the best partition point
and the fastest partition point.
According to the inference time and the speed-up in the table,
we can see that sometimes the speed of collaborative inference is faster
than that of the cloud inference only.
This is due to the large transmission overhead in the low-bandwidth wireless environments.
In collaborative inference,
we only need to download the parameters required by the edge inference, which can significantly reduce the size of download data.
If users need to achieve the fastest inference speed, the fastest partition point should be selected.
If users need to avoid privacy disclosure, the best partition point should be selected.
In addition, quantized neural networks do not lead to a significant drop in accuracy (usually less than 1\%).

Fig. \ref{p_results} shows the collaborative inference time of each candidate layer in the wireless network environments.
We take AlexNet as an example.
Each bar represents a network partition, which consists of three parts: edge inference, data upload and cloud inference.
After auto-tuning of framework, conv5 layer is selected as the best partition point (marked with a hollow pentagram) and the fastest partition point (marked with a filled pentagram).
On edge devices, we feed input data to the neural network and perform inference of layers from conv1 to conv5.
The output data of conv5 (pool and relu are merged) is uploaded to the cloud,
and then the inference of layers from fc6 to fc8 is executed in the cloud.
The approach of collaborative inference achieves 1.7$\times$ speed-up.
It can be seen that the accuracy drop of the network is trivial, and the largest accuracy loss in all partitions is only -0.11\%.

\section{Related Work}
Recently, many neural network quantization methods have been proposed.
Gong Y. et al. \cite{Gong2014Compressing} and Cheng J. et al. \cite{Cheng2017Quantized} explored scalar and vector quantization methods for compressing DNNs.
Zhou A. et al. \cite{Zhou2017Incremental}, Zhou S. et al. \cite{Zhou2016DoReFa} proposed fixed-point quantization methods.
Cuervo E. et al \cite{Cuervo2010MAUI} and Kang Y. et al. \cite{Kang2017Neurosurgeon} designed frameworks that support collaborative computing of mobile applications.
Their frameworks perform off-line partition for full-precision neural networks, and ours performs on-line partition for mixed-precision neural networks.
Overall, the application of quantization methods in cloud-edge collaborative inference has not been studied yet.
To the best of our knowledge, it is the first attempt to build  framework for cloud-edge collaborative inference of mixed-precision neural networks.

\section{Conclusion}
In this paper, we propose an auto-tuning neural network quantization framework
for collaborative inference.
We analyze the characteristics of network layers
and provide candidate rules to choose reasonable partition points.
The auto-tuning framework helps developers get the most suitable partition of a neural network.
The cloud-edge mode (i.e. collaborative inference) reduces the storage of inference on mobile devices with trivial loss of accuracy and could protect personal information.

\section*{Acknowledgement}

This work is supported by the National Key R\&D Program of China under Grant No.2017YFB0202002, the Science Fund for Creative Research Groups of the National Natural Science Foundation of China under Grant No.61521092 and the Key Program of National Natural Science Foundation of China under Grant Nos.61432018, 61332009, U1736208.

\bibliographystyle{splncs03}
\bibliography{paper}

\begin{thebibliography}{10}
\providecommand{\url}[1]{\texttt{#1}}
\providecommand{\urlprefix}{URL }

\bibitem{google-gemm}
gemmlowp: a small self-contained low-precision gemm library.
  \url{https://github.com/google/gemmlowp}

\bibitem{tensorrt}
\textnormal{NVIDIA TensorRT}. \url{https://developer.nvidia.com/tensorrt}

\bibitem{Cheng2018A}
Cheng, J., Wang, P., Li, G., Hu, Q., Lu, H.: Recent advances in efficient
  computation of deep convolutional neural networks. CoRR  abs/1802.00939,
  1--12 (2018), \url{http://arxiv.org/abs/1802.00939}

\bibitem{Cheng2017Quantized}
Cheng, J., Wu, J., Leng, C., Wang, Y., Hu, Q.: Quantized cnn: A unified
  approach to accelerate and compress convolutional networks. IEEE Transactions
  on Neural Networks \& Learning Systems (99),  1--14 (2017)

\bibitem{Cuervo2010MAUI}
Cuervo, E., Balasubramanian, A., Cho, D.K., Wolman, A., Saroiu, S., Chandra,
  R., Bahl, P.: \textnormal{MAUI}:making smartphones last longer with code
  offload. In: International Conference on Mobile Systems, Applications, and
  Services. pp. 49--62 (2010)

\bibitem{DengDSLL009}
Deng, J., Dong, W., Socher, R., Li, L.J., Li, K., Li, F.F.: Imagenet: A
  large-scale hierarchical image database. In: Computer Vision and Pattern
  Recognition. pp. 248--255. IEEE Computer Society (2009)

\bibitem{Gong2014Compressing}
Gong, Y., Liu, L., Yang, M., Bourdev, L.D.: Compressing deep convolutional
  networks using vector quantization. CoRR  abs/1412.6115,  1--10 (2014),
  \url{http://arxiv.org/abs/1412.6115}

\bibitem{He2015Deep}
He, K., Zhang, X., Ren, S., Sun, J.: Deep residual learning for image
  recognition. In: Computer Vision and Pattern Recognition. pp. 770--778 (2015)

\bibitem{hinton2012deep}
Hinton, G., Deng, L., Yu, D., Dahl, G.E., Mohamed, A.r., Jaitly, N., Senior,
  A., Vanhoucke, V., Nguyen, P., Sainath, T.N., et~al.: Deep neural networks
  for acoustic modeling in speech recognition: The shared views of four
  research groups. IEEE Signal Processing Magazine  29(6),  82--97 (2012)

\bibitem{Jia2014Caffe}
Jia, Y., Shelhamer, E., Donahue, J., Karayev, S., Long, J., Girshick, R.,
  Guadarrama, S., Darrell, T.: Caffe: Convolutional architecture for fast
  feature embedding. In: ACM International Conference on Multimedia. pp.
  675--678 (2014)

\bibitem{Kang2017Neurosurgeon}
Kang, Y., Hauswald, J., Gao, C., Rovinski, A., Mudge, T., Mars, J., Tang, L.:
  Neurosurgeon: Collaborative intelligence between the cloud and mobile edge.
  ACM Sigplan Notices  52(4),  615--629 (2017)

\bibitem{krizhevsky2012imagenet}
Krizhevsky, A., Sutskever, I., Hinton, G.E.: Imagenet classification with deep
  convolutional neural networks. In: Advances in Neural Information Processing
  Systems. pp. 1097--1105 (2012)

\bibitem{Lecun1998Gradient}
Lecun, Y., Bottou, L., Bengio, Y., Haffner, P.: Gradient-based learning applied
  to document recognition. Proceedings of the IEEE  86(11),  2278--2324 (1998)

\bibitem{lecun2015deep}
Lecun, Y., Bengio, Y., Hinton, G.: Deep learning. Nature  521(7553),  436--444
  (2015)

\bibitem{mikolov2013distributed}
Mikolov, T., Sutskever, I., Chen, K., Corrado, G.S., Dean, J.: Distributed
  representations of words and phrases and their compositionality. In: Advances
  in Neural Information Processing Systems. pp. 3111--3119 (2013)

\bibitem{Simonyan2014Very}
Simonyan, K., Zisserman, A.: Very deep convolutional networks for large-scale
  image recognition. CoRR  abs/1409.1556,  1--14 (2014),
  \url{http://arxiv.org/abs/1409.1556}

\bibitem{Szegedy2015Going}
Szegedy, C., Liu, W., Jia, Y., Sermanet, P., Reed, S., Anguelov, D., Erhan, D.,
  Vanhoucke, V., Rabinovich, A.: Going deeper with convolutions. In: Computer
  Vision and Pattern Recognition. pp. 1--9 (2015)

\bibitem{Zhou2017Incremental}
Zhou, A., Yao, A., Guo, Y., Xu, L., Chen, Y.: Incremental network quantization:
  Towards lossless cnns with low-precision weights. CoRR  abs/1702.03044,
  1--14 (2017), \url{http://arxiv.org/abs/1702.03044}

\bibitem{Zhou2016DoReFa}
Zhou, S., Ni, Z., Zhou, X., Wen, H., Wu, Y., Zou, Y.: \textnormal{DoReFa-Net}:
  Training low bitwidth convolutional neural networks with low bitwidth
  gradients. CoRR  abs/1606.06160,  1--13 (2016),
  \url{http://arxiv.org/abs/1606.06160}

\end{thebibliography}

\end{document}